\documentclass[lettersize,journal]{IEEEtran}
\usepackage{amsmath,amsfonts}
\usepackage{algorithmic}
\usepackage{array}
\usepackage[caption=false,font=normalsize,labelfont=sf,textfont=sf]{subfig}
\usepackage{textcomp}
\usepackage{stfloats}
\usepackage{url}
\usepackage{verbatim}
\usepackage{graphicx}
\def\BibTeX{{\rm B\kern-.05em{\sc i\kern-.025em b}\kern-.08em
    T\kern-.1667em\lower.7ex\hbox{E}\kern-.125emX}}
\usepackage{balance}

\begin{document}
\title{ \boldmath A Cryogenic Sapphire Resonator Oscillator with $10^{-16}$ mid-term fractional frequency stability \unboldmath}

\author{Christophe Fluhr$^\nabla$, Beno\^{i}t Dubois$^\nabla$, Claudio E. Calosso$^\flat$, Fran\c cois Vernotte$^\sharp$, Enrico Rubiola$^{\sharp\, \flat}$ and Vincent Giordano$^\sharp$
\thanks{Manuscript created August, 2022}
\thanks{$^\nabla$ France Comt\'{e} Innov, Besan\c{c}on, France.}
\thanks{$^\sharp$ FEMTO-ST Institute, Dept.\ of Time and Frequency, Universit\'{e} de Bourgogne and Franche-Compt\'{e} (UBFC), and Centre National de la Recherche Scientifique(CNRS), E-mail: giordano@femto-st.fr,~~  
Address: ENSMM, 26 Rue de l'Epitaphe, 25000 Besan\c{c}on, France.}
\thanks{$^\flat$  Physics Metrology Division, Istituto Nazionale di Ricerca Metrologica INRiM, Torino, Italy.}}

%\markboth{Journal of \LaTeX\ Class Files,~Vol.~18, No.~9, September~2020}%
%{How to Use the IEEEtran \LaTeX \ Templates}

\maketitle

\begin{abstract}
\boldmath
We report in this letter the outstanding frequency stability performances  of an autonomous cryogenique sapphire oscillator presenting a flicker frequency noise floor below $2\times~10^{-16}$ near $1,000 s$ of integration time and a long term Allan Deviation (ADEV) limited by a random walk process of $\sim 1\times 10^{-18} \sqrt{\tau}$. The frequency stability qualification at this level called for the implementation of sophisticated instrumentation associated with ultra-stable frequency references and ad hoq averaging and correlation methods.
\unboldmath
\end{abstract}

\begin{IEEEkeywords}
Time and frequency metrology, ultra-stable oscillators, frequency stability.
\end{IEEEkeywords}

\section{Introduction}

\IEEEPARstart{T}{ests} of fundamental physic \cite{wolf03,eisele-2009,takamoto-2020,campbell-2021}, radioastronomy \cite{rioja2012,alachkar-2018} or fundamental and applied metrology \cite{dolgovskiy-2014,zobel-2019} make an extensive use of ultra-stable frequency sources, for which there is a constant demand for improved frequency stability performance for measurement time ranging from $1$ to $10^{6}$~s. Atomic frequency standards are of course preferred when accuracy and long-term frequency stability are required. But even in this case, an ultra-stable signal source based on a high Q-factor macroscopic resonator is needed to reach the ultimate frequency stability of the atomic clock \cite{santarelli-1998,abgrall2016,robinson2019}. These secondary references, means that are not based on the observation of an atomic resonance, are built around an ultrasonic quartz resonator for the RF or VHF band, a dielectric resonator for microwave, or a Fabry-Perrot cavity for optics. The macroscopic resonator can be integrated directly in the loop of a self-sustained oscillator, or used as a passive reference on which a flywheel oscillator is stabilized. The high Q-factor and the power-handling capability of the macroscopic resonator guarantee a high short term frequency stability. 
However, at mid and long term, i.e. for integration times ranging to say from $10$~s to few days, the oscillator frequency stability is degraded by the fluctuations of the resonator natural frequency. \\

The design of a signal source with the highest frequency stability in the widest integration time range is challenging. Indeed, we have to manage a great number of perturbation sources impacting the frequency stability at different integration times. The means of overcoming all these disturbances are often contradictory between them, and thus tradeoffs have to be found. For example, increasing the signal power increases the signal to noise ratio and thus is favourable for the short term frequency stability. But it can also induce a resonator non-linearity, which makes the resonant frequency sensitive to the signal amplitude \cite{horton-2006,chang97,jap-2014}, as thus will impact the long term frequency stability. \\

The metrological aspect is also very challenging when we have to optimize and qualify a new type of ultra-stable source. If a better reference is not available, two almost identical units have to be implemented and compared. As it is impossible to ensure that each signal source contributes equally to the observed frequency fluctuations, the measured result gives only an overestimated ADEV. If an improvement is made to one unit, its impact on the measurement result can be hidden by fluctuations of the other source. A more efficient way to get the intrinsic frequency stability of the oscillator to be qualified, is to apply the three-cornered-hat (TCH) method or other equivalent Covariance method \cite{vernotte-2016}. The price to be payed is the need of two other signal sources with comparable performances. 
These methods have actually been used for several types of ultra-stable oscillators \cite{uffc-2016,calosso2019,oelker2019}, providing a better understanding of the main frequency stability limitations. However, the TCH or Covariance methods fall when correlations exist between two of the signal sources that are comparated, giving non realistic variances. One of the major issue comes from mid- or long term environment fluctuations that could induce such correlations.\\

The Cryogenic Sapphire Oscillator (CSO) is an autonmous microwave oscillator able to meet the requirements for many very demanding applications. The first CSO generation incorporating a 6 or 8 kW cryocooler as the cold source, demonstrated  an ADEV $\sigma_{y}(\tau)<1\times 10^{-15}$ for $1$~s$\leq \tau \leq 10,000$~s with $<1\times 10^{-14}$/day drift \cite{rsi-2012,journal-physics-2016}. A second CSO generation, code-named ULISS-2G, consuming only $3$~kW single phase is now commercially available. For these instruments the conservative ADEV specification is:  $\sigma_{y}(\tau)\leq 3\times 10^{-15}$ for $1$~s$\leq \tau \leq 10,000$~s and better than $1\times 10^{-14}$ over one day \cite{cryogenics-2016,www.uliss}.  We already build, validated and delivered five ULISS-2G CSOs to different international metrological Institutes \cite{fluhr-2022-ursi}. The sixth unit has been operating for the first time in March 2022 and, at the time of writing, is still under validation process. Although its design is identical to the previous machines, this last unit showed improved performances from the first tests.\\

In this paper, we report on the frequency stability characterization of this new CSO  between $1~$s to about $3$ days. Its ADEV is below $2\times 10^{-16}$ between $100$ to $10^{4}$ s.

%%%%%%%%%%%%%%%%%%%%%%%%%%%%%%
\section{The CSO under test}
%%%%%%%%%%%%%%%%%%%%%%%%%%%%%%
\noindent The design of this new CSO, code-named U10, is described in detail in \cite{cryogenics-2016}. It incorporated a $3$~kW pulse-tube cryocooler to cool near the liquid helium temperature a 54-mm-diameter and 30-mm-height resonator machined in a high purity sapphire monocrystal. This resonator operates on the quasi-transverse magnetic whispering-gallery mode WGH$_{15,0,0}$ at $\nu_{0}= 9.99$~GHz. The sapphire resonator frequency shows a turnover temperature $T_{0}=6.2$~K for which the resonator sensitivity to temperature variations nulls at first order. The appearance of this turning point results from the presence in Al$_{2}$O$_{3}$ of a small amount of paramagnetic impurities as Cr$^{3+}$ or Mo$^{3+}$ and is specific to each resonator. The CSO is a Pound-Galani oscillator: the resonator is used in transmission mode in a regular oscillator loop, and in reflection mode as the discriminator of the classical Pound servo. The sustaining stage and the control electronics are placed at room temperature. The CSO output signal at the resonator frequency $\nu_{0}$ drives the frequency synthesizer, which eventually delivers several output frequencies: $10$~GHz, $100$~MHz and $10$~MHz in the typical implementation. Eventually the synthesizer outputs can be disciplined at long term on an external 100 MHz signal coming from an Hydrogen Maser for example.\\

For the measurements described here, U10 was implemented in the laboratoy workshop
 equipped only with the standard air-conditioning system of the flat. Depending on the sunlight the temperature near the cryostat can vary of several degrees during the day. Moreover the workshop is in free access for laboratory staff and this makes it impossible to maintain an undisturbed ambient for the duration of the measurement (few days).

\section{Measurement set-up}
\begin{figure}[h!!!!!!!!!!]
\centering
\includegraphics[width=3in]{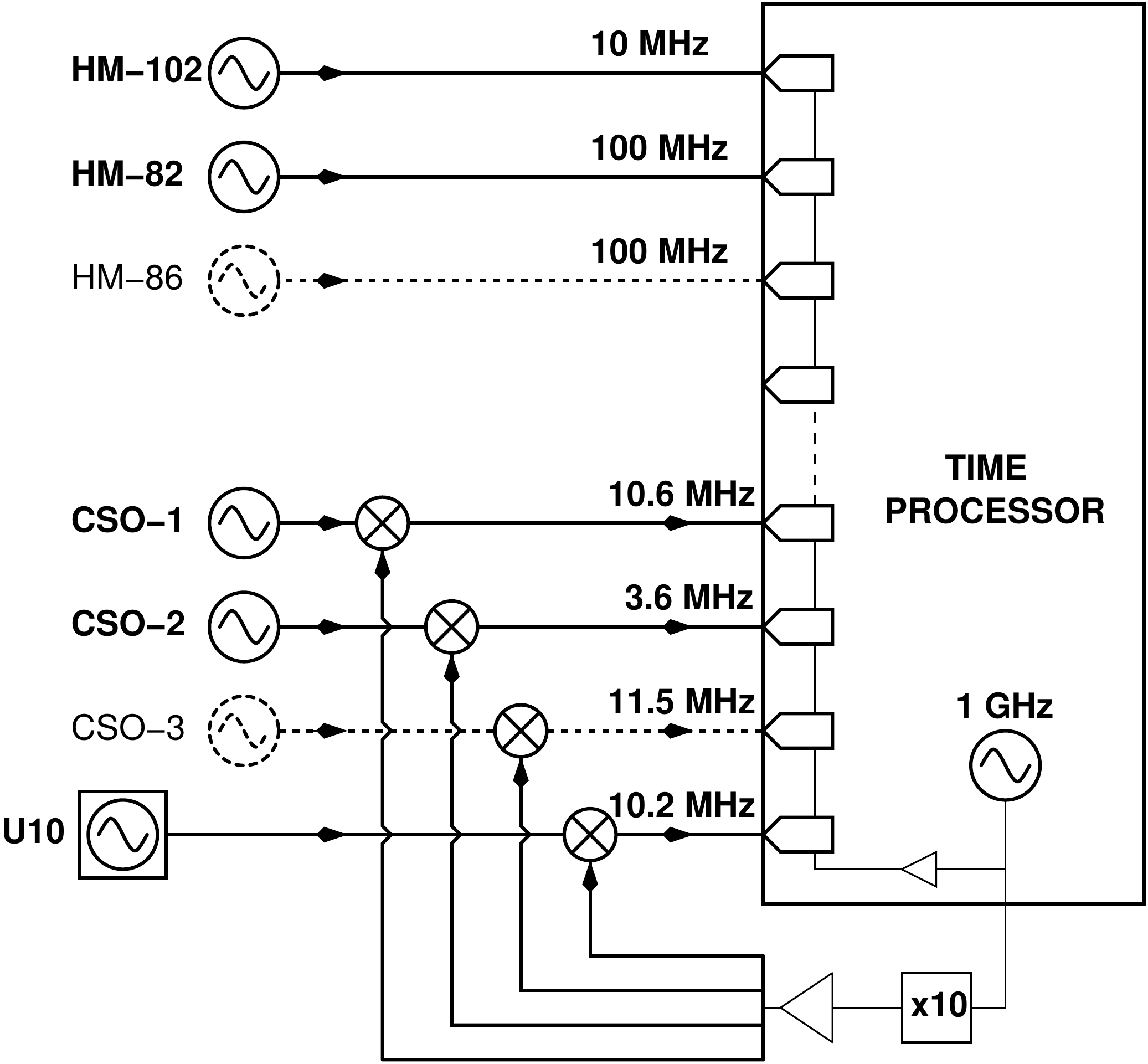}
\caption{Measurement set-up. U10 was compared to the Cryogenic Sapphire Oscillators CSO-1, CSO-2 and the Hydrogen Masers HM-102 and HM-82. The two others reference sources, i.e CSO-3 and HM-86 were used for healthy checks.}
\label{fig1}
\end{figure}

The accurate qualification of U10 between $1$~s to about $3$~days has been made possible by the availability of a multichannel real-time phasemeter designed by one of the authors \cite{calosso-2013-tracking-DDS} . This instrument, i.e. the \it Time Processor\rm, is based on the  Tracking Direct Digital Synthesizer (TDDS) technology. In short, a dedicated DDS is phase-locked to each input signal and the phase information of the input with respect  to the local oscillator is extracted from the phase-control word. The data are normalized to phase time, so that channels at different frequencies can be compared directly. The newly implemented version of the Time Processor is able to compare together up to 16 independant signal sources or beatnotes at different frequency. Each input is characterized by an acquisition and lock range of $5-400$~MHz, and a cut-off frequency ($f_{H}$) of 5 Hz. The one channel resolution in term of Allan Deviation (ADEV) is $\sigma_{y}(\tau)=1.7\times 10^{-14}/\tau~~(2.1\times 10^{-14}/\tau)$ for a $100~(10)$~MHz input carrier. This limitation is set by the intrinsic phase noise of the DDSs. The measurement set-up is schematized in the figure \ref{fig1}. 
 
To perform our measurements, we dispose of the OSCILLATOR-IMP platform RF and microwave ultrastable references: a set of 3 Hydrogen Masers (HM), as well as a set of 3 high-performance CSOs, placed in two independant temperature stabilized room at $22 \pm 0.5$~$^{\circ}$C \cite{www.osc-imp}. The first inputs of the Time Processor receive the Hydrogen Maser signals at $10$~MHz, or $100$~MHz and compares them with the $1$~GHz local oscillator of the instrument. 
The latter is compared also with U10 and with the three reference CSOs by means of a by-10 frequency multiplier and of frequency mixers that produce three beatnotes in the 10 MHz range. The instrument measures the three beatnotes and scales the results to the nominal frequencies of the CSOs. In this way, the residual noise of the instrument is reduced by about three orders of magnitude and becomes completely negligible. In a second step, the phase-time difference of U10 with respect to the other channels is computed and used to calculate the two sample covariance of U10 with respect to two CSOs and two HMs. We point out that these differences cancel out the contribution of the local oscillator that thus does not contribute to the measure. The results presented here have been obtained using CSO-1, CSO-2, IM-102 and IM-82 as references. The permutations done with CSO-3 and IM-86 led to the same results, demonstrating the reproducibility of the procedure.

\section{Results and analysis}

U10 has been turned on for the first time in March, 2022. Then, during the first month, the parameters of the different control loops were adjusted and optimized. During this phase, the CSO experienced significant variations in its operating parameters. Thereafter, the CSO was left running and the first stability assessment can began. The following evaluations were carried out just after this adjustment phase, and the CSO still had a significant drift, i.e. $6\times 10^{-14}/$day that slowly decreases over time. Thereby, for all the results presented here, the ADEV calculations have been computed after a drift removal.\\

At short term, the three reference CSOs are far better than the Hydrogen Masers. They reach an ADEV better than $1\times 10^{-15}$ for $\tau=1$~s, while it is typically $7\times 10^{-14}$ for the HMs. Thanks to correlation and averaging that are inherent to 2 sample covariance, the influence of the reference sources frequency fluctuations on the measured ADEV is reduced by $m^{1/4}$, $m$ being the number of measurements at a given integration time $\tau$. The two CSOs are used for the evaluation of the short-term, since their frequency noise is much lower than masers and, thanks to the number of averages, their contribution is below $1\times 10^{-16}$. Such level of resolution could not be reached by using the two HMs, since it would require an unrealistic acquisition time.

For $\tau \geq 700~$s, the CSOs frequency fluctuations are partially correlated due to an inherent temperature control pumping in the CSO room. The covariance method applied in this integration time range gives for U10 a negative and unrealistic ADEV. Such a level of correlation does not exist between the HMs frequency fluctuations, because i) the HM thermal sensivity is about 10 times lower than those of CSOs, ii) the heat generated by the instruments is lower in the HM room, iii) the situation in the building is more favorable for room HM making the ambient temperature regulation more easy to tune. \\

\begin{figure}[h!!!!!!!!!!]
\centering
\includegraphics[width=\columnwidth]{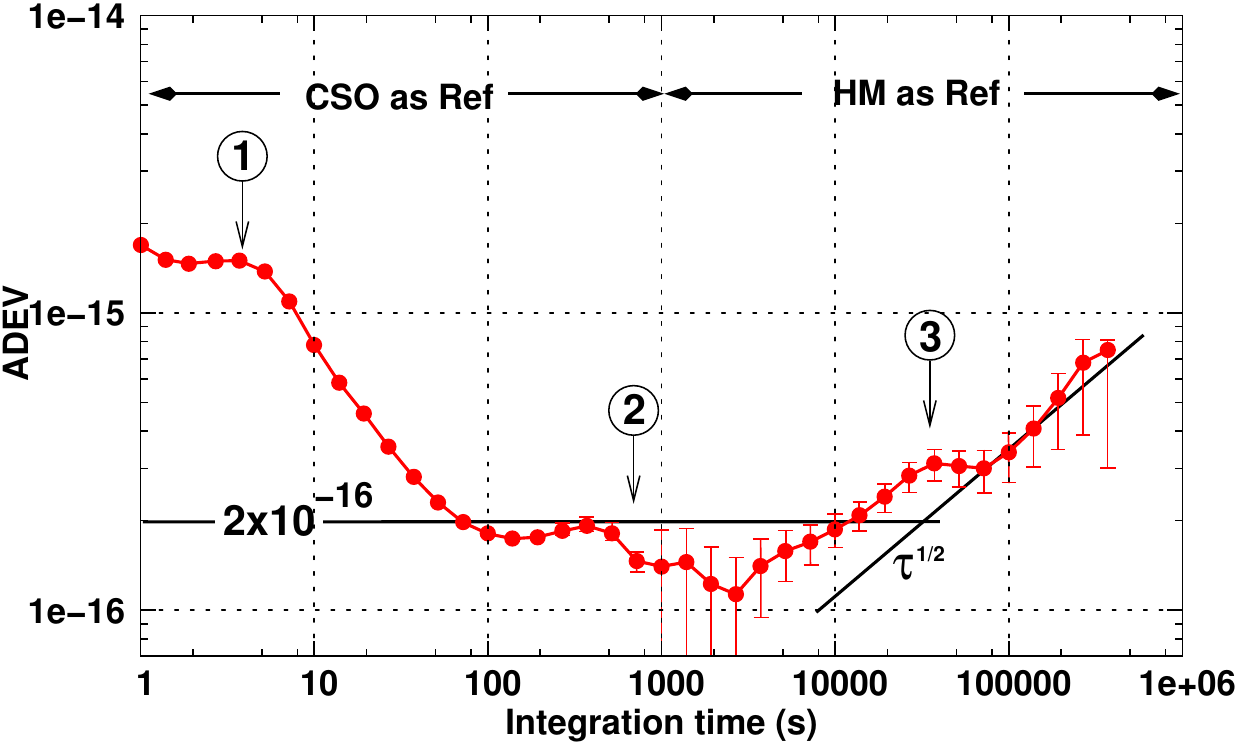}
\caption{U10 ADEV mean estimates. Error bars : $68\%$ confidence intervals.}
\label{fig2}
\end{figure}

Thus, the figure \ref{fig2} represents the U10 ADEV, obtained by combining the calculations made with two different sets of data. Until $\tau=700$~s, the U10 ADEV is determined from the comparison with CSO-1 and CSO-2, and for the longer integration times, the comparison with HM-102 amd HM-82 is used.

The most important result is that the relative frequency instability of U10 is less than $2\times10^{-16}$~for~$100$~s $\leq \tau \leq 10,000~$s, making the CSO the best commercially available oscillator based on a macroscopic resonator.
At longer integration time, the U10 ADEV appears limited by a random walk process such as $\sigma_{y}(\tau) \sim 1.1\times 10^{-18} \sqrt{\tau}$.\\
 Some deviations of the actual ADEV from these two asymptotes can be explained. At short term, the hump $\#1$ around few seconds results from the imperfect resonator temperature stabilization as demonstrated in \cite{uffc-2016,fluhr-2022-ursi}. Then, the resonator and its suroundings thermal mass filter the residual temperature fluctuations, and the ADEV rolls off with a slope $\sim \tau^{-1}$ until about $100$~s. 
The notch  $\#2$, which appears just before $1,000$ s, is the residual of the unrealistic ADEV roll off due to the correlation existing in the CSO references frequency fluctuations. Eventually, at around half a day, the small bump, i.e $\#3$ in the figure \ref{fig2}, can be the signature of the daily temperature fluctuations revealed by the U10 residual sensitivity.

\section{Acknowledgements}

This work is partially supported by (i) ANR, FIRST-TF Network, Grant ANR-10-LABX-48-01, (ii) ANR Oscillator IMP project, Grant ANR11-EQPX-0033-OSC-IMP, (iii) ANR EUR EIPHI Graduate School, Grant ANR-17-EURE-00002, and (iv) grants from the R\'egion Bourgogne Franche Comt\'e intended to support the above projects.

\bibliographystyle{ieeetr}

\end{document}